\documentclass[12pt]{iopart}
\usepackage{graphicx}
\usepackage{color}

\newcommand*\red {\textcolor[rgb]{0.00,0.00,0.00}} %BLACK

\newcommand{\°}{$^{\circ}$}
\newcommand{\º}{$^{\circ}$}

\newcommand{\DIPC}[0]{{
Donostia International Physics Center, 20018 Donostia-San Sebasti\'an, Spain}}
\newcommand{\CFM}[0]{{
Centro de F\'\i sica de Materiales (CSIC-UPV-EHU) and Materials Physics Center (MPC), 20018 San Sebasti\'an, Spain}}

\newcommand{\APLI}[0]{{
Departamento de F\'{\i}sica Aplicada I, Universidad
del Pa\'{\i}s Vasco UPV/EHU, 20018 San Sebasti\'an, Spain}}
\newcommand{\IKER}[0]{{
Ikerbasque, Basque Foundation for Science, 48013 Bilbao, Spain}}

\newcommand{\TUDD}[0]{{
Institut f\"ur Festk\"orper- und Materialphysik, Technische Universit\"at Dresden, D-01062 Dresden, Germany}}

\usepackage{iopams}
\begin{document}

\title[Boron nitride monolayer growth on vicinal Ni(111) substrates]{Boron nitride monolayer growth on vicinal Ni(111) surfaces systematically studied with a curved crystal}

\author{L. Fernandez$^1$, A. A. Makarova$^1$, C. Laubschat$^1$, D. V. Vyalikh$^{2,3}$, D. Yu. Usachov$^4$, J. E. Ortega$^{2,5,6}$, and F. Schiller$^{2,6}$}

\address{$^1$ \TUDD}
\address{$^2$ \DIPC}
\address{$^3$ \IKER}
\address{$^4$ St. Petersburg State University, Physical Department, 7/9 Universitetskaya Nab., St. Petersburg 199034, Russia}
\address{$^5$ \APLI}
\address{$^6$ \CFM}
\ead{frederikmichael.schiller@ehu.es}

%\vspace{10pt}
%\begin{indented}
%\item[]August 2017
%\end{indented}

\begin{abstract}
The structural and electronic properties of hexagonal boron nitride (hBN) grown on stepped Ni surfaces are systematically investigated using a cylindrical Ni crystal as a tunable substrate. Our experiments reveal homogeneous hBN monolayer coating of the entire Ni curved surface, which in turn undergoes an overall faceting. The faceted system is defined by step-free hBN/Ni(111) terraces alternating with strongly tilted hBN/Ni(115) or hBN/Ni(110) nanostripes, depending on whether we have A-type or B-type vicinal surfaces, respectively. Such deep substrate self-organization is explained by both the rigidity of the hBN lattice and the lack of registry with Ni crystal planes in the vicinity of the (111) surface. The analysis of the electronic properties by photoemission and absorption spectroscopies reveal a weaker hBN/Ni interaction in (110)- and (115)-oriented facets, as well as an upward shift of the valence band with respect to the band position at the hBN/Ni(111) terrace.
\end{abstract}

%
% Uncomment for keywords
\vspace{2pc}
\noindent{\it Keywords}: curved crystal, faceting, hBN monolayer, vicinal Ni(111) surface
%
% Uncomment for Submitted to journal title message
%\submitto{\TDM}
%
% Uncomment if a separate title page is required
%\maketitle
%
% For two-column output uncomment the next line and choose [10pt] rather than [12pt] in the \documentclass declaration
%\ioptwocol
%

\section{Introduction}
Since the discovery of graphene, a wide diversity of atomic-layer-thick, two-dimensional (2D) materials with varied properties have emerged.
Of particular interest are those that exhibit semiconducting behavior, such as hexagonal boron nitride (hBN). hBN is isoelectronic to graphene and has also a honeycomb lattice formed by alternating nitrogen and boron atoms~\cite{Pakdel2014_ChemSocRev,Rokuta1997_PRL}, but in contrast to the semi-metallic graphene, its band structure presents a 4-6 eV band gap at the Fermi energy ~\cite{Zunger1976_PRB,SOLOZHENKO2001_JPhysChemSol,Watanabe2004_NatMat,Watanabe2009_NatPhot}. This makes it particularly attractive for applications in microelectronics ~\cite{Rao2013_AngewChem,Zhang2017_JMatChemC}, either alone or in combination with other 2D materials, such as graphene ~\cite{Usachov2010_PRB,Liu2011_NanoLett,Britnell2012_Science,Yang2013_NatMat,Kang2014_AdFuncMat}.
Therefore it exists an increasing demand for large size hBN films that can be transferred to different surfaces, but these films first have to be synthesized on adequate substrates by different growth techniques~\cite{Oh2016_NpgAsiaMat,Wang2015_AdvMat,Meng2017_Small}.
Additionally, hBN is structurally robust and chemically inert, properties that are particularly relevant for coating~\cite{Li2014_Nanotechnology,Caneva2017_ACS_ApplMat} and hydrogen storage~\cite{Spaeth2017_2DMat} applications.

The increasing interest on single-layer hBN, doped and intercalated hBN/metal interfaces, and low-dimensional hBN structures, such as nanoribbons, has prompted the search of appropriate substrates, in order to achieve functional hBN layers and nanostripes with abrupt interfaces. The growth challenge for hBN is similar to the one faced by other hetero-epitaxial systems in the past~\cite{Schukin2013,Markov2017}, and hence one can gain from the knowledge acquired over decades, particularly about the balance and hierarchy of driving forces that govern crystal growth. Among them, atomic lattice matching, which is often sought to prevent large accumulation of stress in extended films, and nano-templating, as a way to optimize growth by altering energetics and kinetics. In this context, stepped (or vicinal) surfaces emerge as simple self-organized growth templates, known to facilitate the synthesis of one-dimensional (1D) nanostructures, such as atomic rows and nanostripes~\cite{Schukin2013} and to promote single rotational domains in growing films~\cite{Kuntze2002_APL,Liu2003_Nanotech,Wang2011_ACSNano}.
%, and to modulate substrate registry via the step spacing~\cite{Bachmann2003_SS,Riemann2005_PRB}.

Despite the advantages and versatility of vicinal surfaces as growth templates, very little has been done in regard of 2D, monolayer-thick materials~\cite{ROKUTA1999_SS,Usachov2007_SSP,Srut2013_Carbon,Kajiwara2013_PRB,Celis2018_PRB}. One obvious reason is the added structural complexity of the stepped interface. Additionally, vicinal surfaces are not rigid frames that remain static or passive during epitaxy, but, on the contrary, they often respond to overlayer growth by step-edge roughening, step bunching or nano- and mesoscale faceting~\cite{Bachmann2003_SS,Riemann2005_PRB,Johnson1994_PRL,Hornvonhoegen1999_SS}. All this diversity of phenomena is difficult to predict and, given the infinite variety of vicinal planes for each crystal orientation, their practical understanding can gain from systematic investigations using curved surfaces. In this context, it has recently been shown that a very facile, and hence effective implementation of the curved surface approach is achieved with cylindrical sections of single crystals~\cite{Ortega2011_PRB,Walter2015_NatComm,Miccio2016_NL,Ortega2018_NJP}, such as the one sketched in Fig.~\ref{STM_LEED}. This represents a $\alpha=\pm 15^{\circ}$ portion of a cylindrical Ni single crystal, with axis parallel to the [1$\bar 1$0] direction, and surface center oriented along the [111] direction. Therefore, almost the entire family of Ni(111) vicinal planes with A-type (\{100\} microfacet) and B-type (\{111\} microfacet) close-packed steps are found at both sides of the (111) center, up to the (335) and (221) crystal directions.

Here we thoroughly explore the synthesis of a single layer hBN on vicinal Ni(111) surfaces with close-packed steps using the curved sample [in short, c-Ni(111)] described in Fig.~\ref{STM_LEED}. The c-Ni(111) crystal has the advantage of having the Ni(111) surface at the center of the sample, and hence we can use the well-studied hBN/Ni(111) interface as a reference~\cite{Rokuta1997_PRL,Gamou1997_SR_RITU,Auwaerter1999_SS,AUWARTER2004_ChemMat}. Previous studies using Ni foils with differently oriented grains already suggest different growth scenarios~\cite{Lee2012_RCSAdv}.
Beyond Ni, the synthesis of the hBN monolayer has been investigated at many high-symmetry metal surfaces \cite{Rokuta1997_PRL,Gamou1997_SR_RITU,Auwaerter1999_SS,AUWARTER2004_ChemMat,Lee2012_RCSAdv,Usachov2018_PRB,Corso2004_Science, PREOBRAJENSKI2007_CPL,Goriachko2007_Langmuir,Martoccia2008_PRL,Orlando2014_ACSNano,Farwick2016_ACSNano, PETROVIC2017_ASS,Morscher2006_SS,PAFFETT1990_SS,PREOBRAJENSKI2007_PRB,CAVAR2008_SS,PREOBRAJENSKI2005_SS,
Joshi2012_NanoLett,Mueller2010_PRB,Greber2006_EJSSN,Vinogradov2012_Langmuir,Mueller2008_SS,HERRMANN2018_SS,CORSO2005_SS, Allan2007_NanoScReLe}. It is generally observed that electronic and structural properties of the epitaxial hBN monolayer are influenced by both the strength of the interface chemical bond and the registry, that is, the lattice-matching with the substrate. At close-packed surfaces, the interaction varies from strong hBN chemisorption on Ni(111), Co(0001)~\cite{Usachov2018_PRB}, Rh(111)~\cite{Corso2004_Science}, Ru(0001)~\cite{PREOBRAJENSKI2007_CPL,Goriachko2007_Langmuir,Martoccia2008_PRL} and Ir(111)~\cite{Orlando2014_ACSNano,Farwick2016_ACSNano,PETROVIC2017_ASS},  to weak interaction on quasi-noble noble Pd~\cite{Morscher2006_SS} and Pt~\cite{PAFFETT1990_SS,PREOBRAJENSKI2007_PRB,CAVAR2008_SS}, and noble Cu(111)~\cite{PREOBRAJENSKI2005_SS,Joshi2012_NanoLett} and Ag(111)~\cite{Mueller2010_PRB} surfaces. Experiments at (110) surfaces of Ni~\cite{Greber2006_EJSSN}, Fe~\cite{Vinogradov2012_Langmuir}, Cr~\cite{Mueller2008_SS}, Cu~\cite{HERRMANN2018_SS}, Pd~\cite{CORSO2005_SS}, and Mo~\cite{Allan2007_NanoScReLe} reveal the little influence of the crystal orientation on the hBN/metal interaction strength. With respect to the hBN layer structure, the relevant feature is the presence of a characteristic nanoscale corrugation, called ``nanomesh'', in strongly interacting interfaces with poor lattice matching, whereas nearly flat hBN films are observed otherwise. The latter is the case of Ni(111).

Our experiments on the c-Ni(111) curved sample reveal homogenous coating of the Ni substrate, although step-bunching and faceting at all Ni(111) vicinal planes. In the A step-type side of the crystal, we observe a two-phase, hill-and-valley structure, driven by the excellent registry of hBN with flat Ni(111) terraces (first phase) and stepped Ni(115) plane (second phase). In the B step-type side, Ni(111) terraces are also present, but the lack of reasonably matched B-type vicinal planes explains the presence of the incommensurate (5.15$\times$1) hBN/Ni(110) facet, for which a minimum expansion/contraction of the rigid hBN lattice is required. All spectroscopic data in faceted regions are explained by the incoherent superposition of the signal from the (111) terrace and the faceted phase. This allows us a direct comparison of the strength of chemical interactions and other electronic properties, such as the valence band maximum, among the three hBN-covered Ni(111), Ni(115) and Ni(110) interfaces. The differences found are consistent with the variable atomic-scale roughness at each interface, from the flat and sharp hBN/Ni(111), to the rough, stepped (115) substrate, and to the corrugated hBN layer on Ni(110).

\section{Experimental details}
The Scanning Tunneling Microscopy (STM), Low Energy Electron Diffraction (LEED) and Angle-resolved Photoemission Spectroscopy (ARPES) experiments were performed at the Material Physics Center of San Sebastian (Spain), while the X-ray absorption and X-ray photoelectron spectroscopy experiments were carried out at the RGBL beamline at BESSY synchrotron in Berlin (Germany). Both setups were equipped with Low Energy Electron Diffraction (LEED) to check for the ordering of the hBN layer and ensure that the same crystal structure is achieved. The Ni(111) crystal was prepared by repeated cycles of sputtering (1000 V, incidence angle 45\° parallel to the steps) and annealing (750$^{\circ}$C). Previous to the last sputtering/annealing cycle the sample was heated at 10$^{-8}$ mbar O$_2$ for 10 minutes at the same temperature to remove possible carbon impurities from the surface. The hBN monolayer was grown by chemical vapor deposition (CVD). For this purpose, the sample held at 750\° C was exposed to Borazine (B$_3$N$_3$H$_6$, 99\%, Katchem Ltd.) vapor during 10 min at 10$^{-7}$ mbar.

{\bf{Low Energy Electron Diffraction.}}
LEED experiments were performed with a standard three grid setup (Omicron). LEED maps are elaborated with 25 individual patterns, acquired by scanning the electron beam (0.5 mm) at normal incidence with respect to the (111) surface across the curved surface. Energy-dependent measurements were carried out by placing the sample normal to the (111) direction at positions corresponding to vicinal angles of $\pm$12.5\°, and scanning the kinetic energy from 50 to 120 eV in steps of 2 eV.

{\bf{Scanning Tunneling Microscopy.}}

STM measurements were carried out at room temperature using an Omicron VT-STM microscope in the constant current mode. In order to obtain information about the step size distribution depending on the vicinal angle, multiple measurements were performed in areas located along the curvature of the nickel crystal. The scanning direction was usually perpendicular to the step edges.

{\bf{Angle-resolved photoemission spectroscopy.}}
ARPES experiments were performed using a monochromatized Helium discharge lamp (Specs VUV300 and TMM302) at h$\nu$=40.8 eV (He II) and a SPECS Phoibos 150 electron analyzer, with energy and angular resolution set to 200 meV and 1\°, respectively. The sample was held at 150 K during measurements. Intensity distribution maps are composed of 40 channelplate images.

{\bf{X-ray Photoemission and Absorption Spectroscopies}}

XPS spectra were acquired with a hemispherical SPECS Phoibos 150 electron energy analyzer. Measurements were carried out in the normal emission geometry, with photon energies of 270 eV and 480 eV for B 1s and N 1s core-level spectra, respectively, in such a way similar kinetic energy ranges were explored. For energy calibration all spectra were aligned relative to the Au 4$f_{7/2}$ peak of a reference gold sample, set to 84.0 eV binding energy. NEXAFS spectra were recorded in total-electron yield mode.  The drain current from the sample was measured using a Keithley 6517 electrometer. The angle between the sample normal and the photon beam was set to 55\º. All the measurements were carried out at 300 K.

\section{Results and Discussion}

\subsection{hBN monolayer growth and structure}
The growth and structure of the hBN/c-Ni(111) system is analyzed by Scanning Tunneling Microscopy (STM) and Low Energy Electron Diffraction (LEED). Both techniques reveal the homogeneous wetting of the entire substrate by a single hBN monolayer, which in turn induces a deep structural transformation of the curved Ni substrate underneath.
\begin{figure}[tb!]
\centerline{\includegraphics[width=140mm,angle=0,clip]{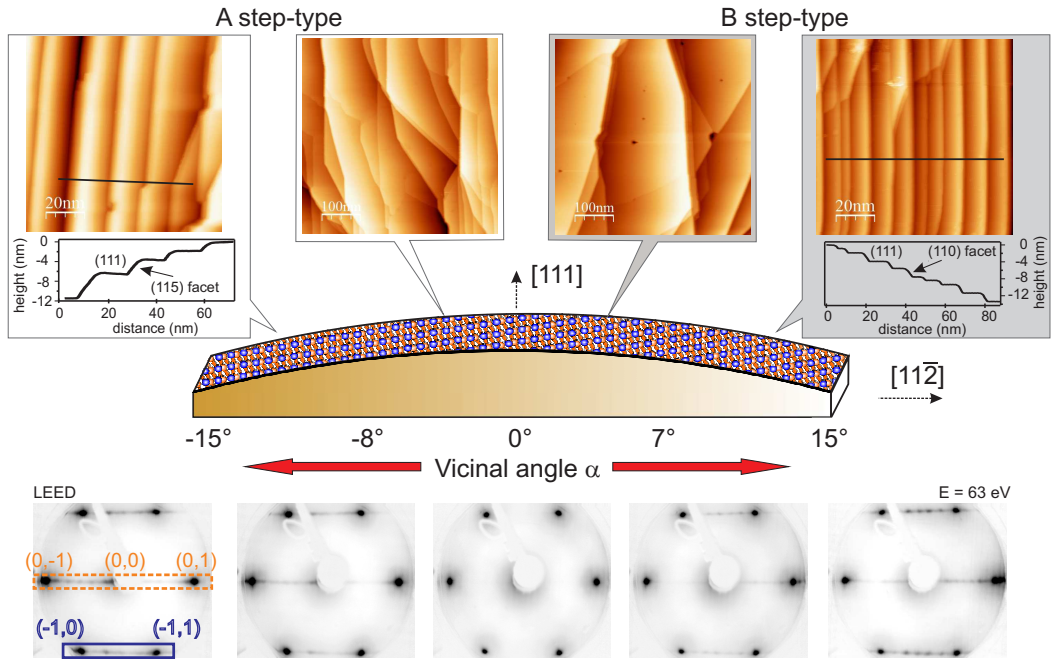}}
\caption{\textbf{hBN monolayer on curved Ni(111)}. Top, Scanning Tunneling Microscopy (STM) and, bottom, Low Energy Electron Diffraction (LEED) for a monolayer of hBN homogeneously covering the curved Ni crystal sketched in the center. The STM images have been taken at the positions roughly indicated over the sample. LEED patterns correspond to the center [(111) plane)], midway (vicinal angle $\alpha=\pm 7^{\circ}$), and densely stepped edges ($\alpha =\pm 15^{\circ}$) of the sample and have been acquired using 63 eV electron impinging parallel to the [111] direction in all cases. Insets in STM images belong to the indicated line profiles, which prove the presence of hBN-covered microfacets.}
\label{STM_LEED}
\end{figure}
The bare c-Ni(111) surface is characterized by patterns of one- and two-atom-high steps~\cite{Ilyn2016_JPCC}, but after hBN growth such simple step lattices transform into hill-and-valley structures of hBN-covered (111) terraces alternated with nanofacets oriented along other Ni crystal directions. Early studies of hBN grown on Ni(755) have already suggested substrate faceting~\cite{ROKUTA1999_SS} and here we confirm that the hBN monolayer growth leads to general faceting at all A-type and B-type vicinal planes. Fig.~\ref{STM_LEED} shows characteristic STM topographies (top) and LEED patterns (bottom) acquired at different positions on the curved sample. The sequence of STM images illustrates the evolution of the faceted surface at the nano- and mesoscale (see more images in the Supplementary Information file). Near the center of the sample one can observe wide (111) terraces and random step bunching, whereas away from the center there is a gradual reduction of the area covered by hBN/Ni(111) terraces, with a simultaneous widening of step-bunched stripes, which become straight and well-oriented 2-10 nm wide crystal facets. Side facets are usually wider on B steps as on A steps for the same vicinal angle of the substrate, e.g., for $\left|\alpha\right|$=10\° on A steps 3 nm wide facets are abundant while on B steps the average facet width is 9 nm. In general the farther from the (111) position, the wider the tilted facet becomes, (see SI for more details). At this range, one can readily determine the local angle of such stepped facets with respect to the (111) reference plane. As indicated in the line profiles, facet inclination angles are (35$\pm$3)$^{\circ}$ at the A-side and (32$\pm$3)$^{\circ}$ at the B-side, which correspond to crystal orientations close to the [115] and [110] directions, respectively. A detailed description of the STM analysis is given in the supplementary information.

By means of LEED we analyze the atomic scale structure of the hBN/c-Ni(111) interface at different sample positions. In the bottom of Fig.~\ref{STM_LEED} we show the characteristic LEED images obtained at the center, at midway and at the edge of the sample. The electron energy $E_p$=63 eV has been chosen to optimize anti-phase interference conditions at the stepped facets, namely intense and well-resolved split-spots at the edges of the curved crystal. The beam incidence is kept perpendicular to the (111) terrace plane during the LEED scan, such that different vicinal planes are probed by a simple horizontal shift of the c-Ni(111) sample. In the center of the sample, the sharp hexagonal pattern corresponds to the hBN monolayer, in registry with the Ni(111) surface plane \cite{AUWARTER2004_ChemMat}. As we move away from the center, second order spots progressively emerge in between the main diffracted beams and along the [11$\bar 2$] direction, that is, perpendicular to surface steps. The intensity evolution across the curved surface indicates that these satellite spots belong to the step bunches and straight facets observed in the STM images.

\begin{figure}[tb!]
\centerline{\includegraphics[width=150mm,angle=0,clip]{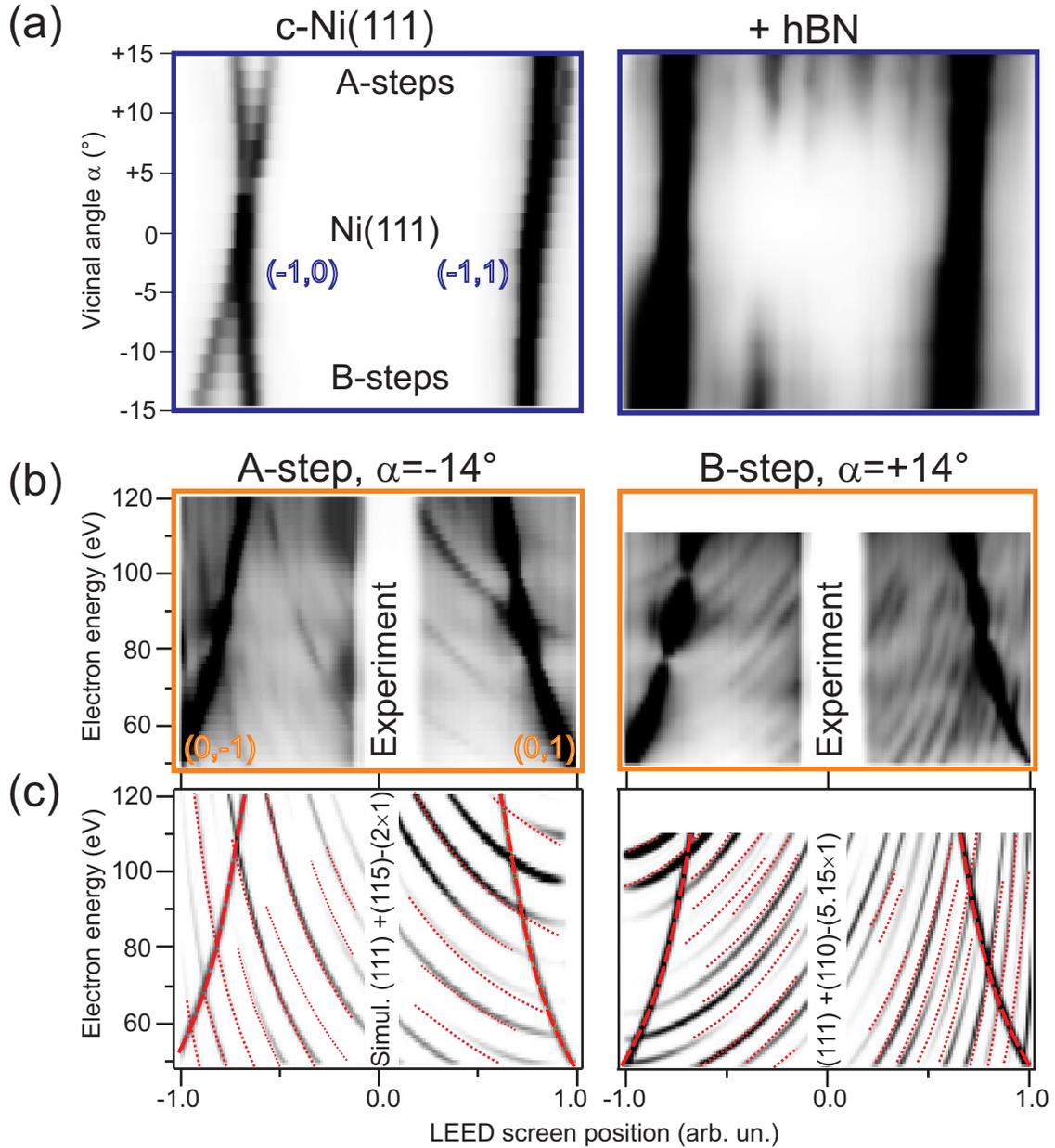}}
\caption{\textbf{LEED analysis of the hBN/c-Ni(111) monolayer interface.} (a) LEED intensity scan across the curved surface (vertical scale) and along the (-1,0)-(-1,1) profile (marked in Fig.~\ref{STM_LEED}) for the clean (left) and  hBN-covered (right) c-Ni(111) sample (electron energy $E_p$=63 eV). The two straight crossing lines in the clean sample indicate homogeneous step arrays. The  vertical streaks in the hBN-covered sample reveal the presence of facets at both A and B sides. (b) $E_p$-dependent LEED plots, elaborated by stacking the $E_p$-dependent (0,-1)-(0,0)-(0,1) profiles (marked in Fig.~\ref{STM_LEED} with a dashed rectangle), and measured at $\alpha=\pm 14^{\circ}$ at the A (left) and B (right) edges of the sample, respectively. Panels (c) show (0,-1)-(0,0)-(0,1) LEED profiles calculated, within the kinematic theory, for the indicated hBN-covered Ni crystal facets. Dashed [(111) facet] and dotted [side facets] lines are the experimental lines extracted from panel (b), which serve to compare with theory.}
\label{LEED}
\end{figure}

A more precise assessment of the evolution of the the hBN/c-Ni(111) system as a function of the vicinal angle can be obtained through the $\alpha$-dependent LEED intensity plots shown in Fig.~\ref{LEED}(a). The images are made by stacking individual intensity profiles along the (-1,0)-(-1,1) line (marked by a solid rectangle in Fig.~\ref{STM_LEED}), which are extracted from 25 LEED patterns sequentially taken across the clean (left panel) and the hBN covered (right panel) c-Ni(111) sample. At the clean surface, and particularly in the (-1,0) spot, we observe a smoothly increasing spot splitting away from the (111) direction that defines two straight crossing lines. These simply reflect the linear dependence of the step spacing expected for vicinal surfaces in a cylindrical geometry \cite{Walter2015_NatComm,Ilyn2016_JPCC}. In contrast, at the hBN/c-Ni(111) interface, the satellite spots between (-1,0) and (-1,1) diffraction rods appear as vertical streaks, with their intensity steadily increasing from the center and toward the two edges of the sample. Since the bulk Ni crystal direction is not modified during the LEED scan, these vertical streaks originate from hBN-covered facets, which are tilted from the (111) direction with a characteristic angle, that is, facets have a single crystal orientation in each side of the crystal.

To determine the atomic structure of the hBN-covered facets that develop at vicinal planes we analyze electron-energy-dependent LEED patterns at the two edges of the c-Ni(111) sample. The images in Fig.~\ref{LEED}(b) are stacks of (0,-1)-(0,0)-(0,1) horizontal line profiles (dotted-line box in Fig.~\ref{STM_LEED}) as a function of the electron energy, taken at $\alpha = \pm 14^{\circ}$. The energy dependence of individual spots defines two sets of ``reflection lines'' in Fig.~\ref{LEED}(b), one for each faceting phase. Reflection lines from the main (0,-1) and (0,1) spots disperse towards the center of the image as the energy is increased, as expected for normal incidence electrons diffracted out of a (1$\times$1) hBN-covered Ni(111) terraces. The remaining lines behave in a different way, since they disperse to one side of the image at increasing energies. They belong to beams diffracted at nanofacet planes tilted away from the (111) surface, and hence contain the atomic scale information of the corresponding hBN/Ni interfaces, that is, the crystallographic orientation of the substrate plane and the registry of the hBN monolayer on top. To deduce both, we simulate the electron diffraction pattern from a two dimensional hBN lattice with different tilting angles (facet orientations) and ($n\times$1) substrate registry within the kinematic approach, and compare the model with the LEED plots of Fig.~\ref{LEED}(b). In the kinematic theory the amplitude of scattering by a wave vector $\Delta \vec{k}$ is given by:
\begin{equation}
	A_{\Delta \vec{k}} = \sum_{j} f_j \exp (i \Delta \vec{k} \vec{R_j}),
\end{equation}
where the sum is taken over all atomic sites $\vec{R_j}$ (in a finite model cluster), and the atomic scattering factors $f_j$ are assumed to be equal for B and N atoms. Into a first approach, a lattice-matched ($n\times$1)hBN/Ni superstructure can be accounted for with a sine-like corrugation of the BN layer with a peak-to-peak amplitude of approx. 0.8 {\AA} and period determined as $ab/|a-b|$, where $a$ and $b$ are the periods of Ni(111) and hBN in the direction perpendicular to steps. The best agreements with the experiments of Fig.~\ref{LEED}(b) are displayed in Fig.~\ref{LEED}(c). For a better comparison, the experimental reflections are also included in Fig.~\ref{LEED}(c) with red dashed [(111) terraces] and red dotted (tilted facets) lines. Note that extinction of some reflections may occur at certain energies due to multiple scattering effects, which are not accounted for in the kinematical approach. Despite this limitation, both the number of diffraction lines and their dispersion are excellently reproduced by the model and for the indicated tilted facet planes and reconstructions, namely the (3$\times$1)-hBN on a (2$\times$1)-Ni(115) at the A side, and the (5.15$\times$1)-Ni(110) incommensurate superstructure at the B side. The Ni(115) and Ni(110) orientations of the tilted facets are in well agreement with the above mentioned STM observations, whereas other registry structures of the hBN layer are discarded (see SI for a deeper LEED analysis of this point). The latter observation is particularly relevant in the case of the Ni(110) facet, since a variety of hBN structures have been found in the corresponding, infinite Ni(110) plane~\cite{Greber2006_EJSSN}. This suggests that registry phases could be eliminated in the presence of steps, that is, with vicinal Ni(110) substrates.

\begin{figure}[tb!]
\centerline{\includegraphics[width=150mm,angle=0,clip]{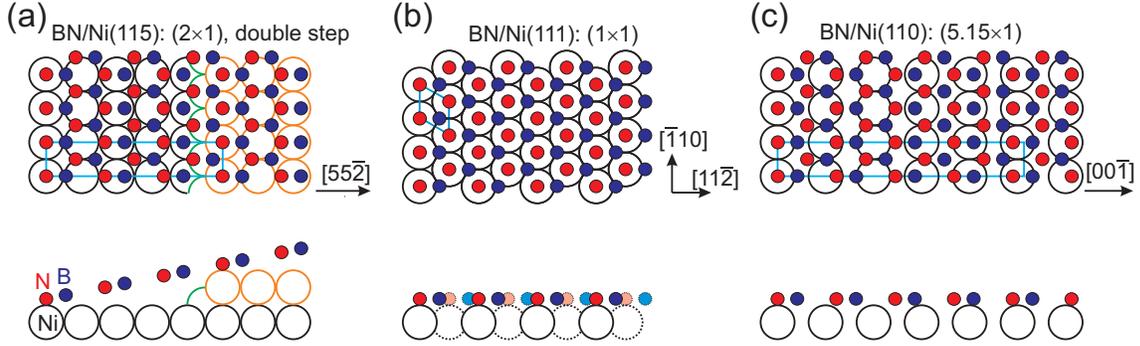}}
\caption{Structural models of hBN on surfaces vicinal to Ni(111), in agreement with STM and LEED results: (a) registry (3$\times$1)-hBN on Ni(115) with double steps observed in A-type vicinals, (b) (1$\times$1)-hBN on Ni(111), and (c) incommensurate (5.15$\times$1) Ni(110) observed in B-type vicinals. In the side view, shaded atoms belong to the next row of atoms behind.}
\label{struc_model}
\end{figure}

Faceting is a two-phase segregation process that is driven by elastic free energy minimization~\cite{Schukin2013}. In the present hBN/c-Ni(111) system, the two coexisting phases are hBN/Ni(111) terraces, on the one side, and hBN/Ni(115) or hBN/Ni(110) facets at A- and B-type steps of the crystal on the other side, respectively. Although phase-boundary and phase-phase interaction energies are relevant, equilibrium phases are generally defined by the interface energy, such that facets exhibiting the lowest atomic mismatch at the interface are, in principle, privileged. The well-matched hBN/Ni(111) phase is naturally expected from the close proximity of surface lattice constants of hBN (2.504 \AA)~\cite{Lynch1966_JCP} and interatomic atom distance at Ni(111) (2.491 \AA)~\cite{BANDYOPADHYAY1977_Cryo}. In fact, hBN grows slightly compressed (0.5\%) on Ni(111), with N atoms occupying Ni on-top sites and B atoms sitting in hollow positions~\cite{Rokuta1997_PRL,Gamou1997_SR_RITU,Auwaerter1999_SS,Grad2003_PRB,Sun2014_JAP,EBNONNASIR2015_SRL}, as depicted in the center of Fig.~\ref{struc_model}. For surfaces vicinal to the (111) direction, we expect the spontaneous bunching of steps forming microfacets with optimal hBN/Ni matching. In the Supporting Information file we evaluate the degree of hBN mismatch to all Ni crystal planes from the (111) to the (001) and the (110) surfaces, in the A and B side of the curved crystal, respectively. In the A side one finds good matching at 38.5$^{\circ}$ away from the (111) plane, namely at the (115) direction \cite{note1}. Here the hBN/Ni(115) registry [(3$\times$1)-hBN on (2$\times$1)- Ni(115)] requires 0.5\% compression of the hBN lattice, similar to that of the (111) surface. In contrast, other Ni planes in the vicinity of the (111) surface require a large, unphysical strain of the very rigid hBN lattice, in order to achieve lattice-matching (see tables in the SI). Notably, both (2$\times$1) and (1$\times$1)-Ni(115) superstructures are equally matched (see the SI), but the latter leads to an non-rectangular LEED pattern that is not observed. As proposed in the left side of Fig.~\ref{struc_model}, the (2$\times$1) pattern is likely due to terrace-width/step-height doubling within the Ni(115) facet plane, which does not change the vicinal angle, nor the hBN lattice strain, but correctly explains the LEED results \cite{note2}. This preference for double-steps at the Ni side of the hBN/Ni(115) interface could be related with higher order contributions to the system free energy, such as the step energy itself (double versus single), or the step-step elastic interaction, which decreases as the step spacing increases~\cite{Ilyn2016_JPCC}.

On the B-side of the c-Ni(111) surface the best kinematic LEED simulation is found for an incommensurate hBN layer covering Ni(110) facets. The solid lines in Fig.~\ref{LEED}(c) have been obtained assuming a (5.15$\times$1) modulation superimposed to the hBN lattice. As sketched in the right side of Fig.~\ref{struc_model}, the (5.15$\times$1) Ni periodicity along the [1$\bar 1$2] direction results from the superposition of a 0.8\%-stretched (4.15$\times$1) hBN layer. In reality, there is a close proximity with the commensurate (4$\times$1)-hBN-on-(5$\times$1)-Ni(110), but a notorious difference with the (6$\times$1) and (5$\times$7) registry structures reported for hBN on Ni(110) ~\cite{Greber2006_EJSSN}.
A (5$\times$7) superstructure can be excluded because it would produce extra spots between the line like spot structures. Based on a detailed LEED analysis, also the proposed (6$\times$1) periodicity is discarded in the present hBN/Ni(110) nano-sized facets (see SI).
Contrary to A-type vicinals, a reasonable hBN/Ni lattice matching cannot be found across the entire family of B-type crystal planes, namely from the (111) surface and up to the (110) (see SI). However, from the point of view of the hBN elastic energy, there is a clear advantage associated with the (5.15$\times$1) one-dimensional incommensurability. It is important to remind that an extra compression is still needed to keep the hBN/Ni registry parallel to the steps (along the [1$\bar 1$0] direction), such that the 0.8\% stretching along [1$\bar 1$2] appears as a compensating effect. In fact, in terms of atomic density variation $\Delta \rho$ with respect to the pristine hBN monolayer, $\Delta \rho$ of the incommensurate hBN layer of Fig.~\ref{LEED}(c) is much smaller ($\Delta \rho$=+0.3\%) compared to that of hBN/Ni(111) ($\Delta \rho$=-1.0\%) or hBN/Ni(115) ($\Delta \rho$=-1.0\%) interfaces. Therefore, although hBN/Ni lattice registry along the [11-2] direction appears to be the driving force to explain faceting in hBN/c-Ni(111) interfaces, an excessive lattice strain (or atomic density variation beyond $\sim \pm$1\%) may force incommensurate structures (or moir\'es). At this point a theoretical analysis is encouraged.

On the Ni(111) surface the nitrogen atom always occupy on-top
positions. Nevertheless, there exists two possibilities for the hollow
positions of the boron atom, namely fcc or hcp sites of the Ni substrate.
Theory found a preference for the fcc sites but only with a very small
energy difference of about 10 meV with respect to the hcp ones
~\cite{Grad2003_PRB,Sun2014_JAP,Che2005_PRB,Huda2006_PRB,Joshi2013_PRB}.
Experiments suggested that both configurations {\red{may}} coexist on the
surface~\cite{AUWARTER2003_SS,AUWARTER2004_ChemMat} but the relative
amount depends on the substrate preparation and the
quality of the molecular precursor. The coalescence of the
triangular islands with B atoms in fcc and hcp sites gives
rise to defect lines, whose density can trigger some of the properties
of the film~\cite{AUWARTER2003_SS}. Being able of tuning the relative
concentration of hcp and fcc domains  would be therefore desired in order
to avoid or enhance metal growth on defect lines~\cite{Osterwalder2003_EJSSN}.
The here investigated intensity profiles of the LEED spots may also
allow one to distinguish the possible existence of single hBN domains. Preferred
hcp or fcc domains can in fact explain the reversed intensity distribution of
the (0,-1) and (0,1) spots in Fig.~\ref{LEED}(b) at A and B
sides, which suggest that hcp and fcc configurations dominate at opposite
sides. Nevertheless other preparations carried out afterwards reveal
no differences in the structure of the hBN/Ni(111) interface on A and B
sides. This suggests that under certain conditions like sample history,
borazine quality or details in the preparation
process, dominant domains can be
achieved. Further experiments are necessary to find the optimal
conditions to achieve single domains. Additionally, LEED I-V theory
would be needed to distinguish fcc or hcp domains.

\subsection{Electronic structure}

\begin{figure}[tb!]
\centerline{\includegraphics[width=130mm,angle=0,clip]{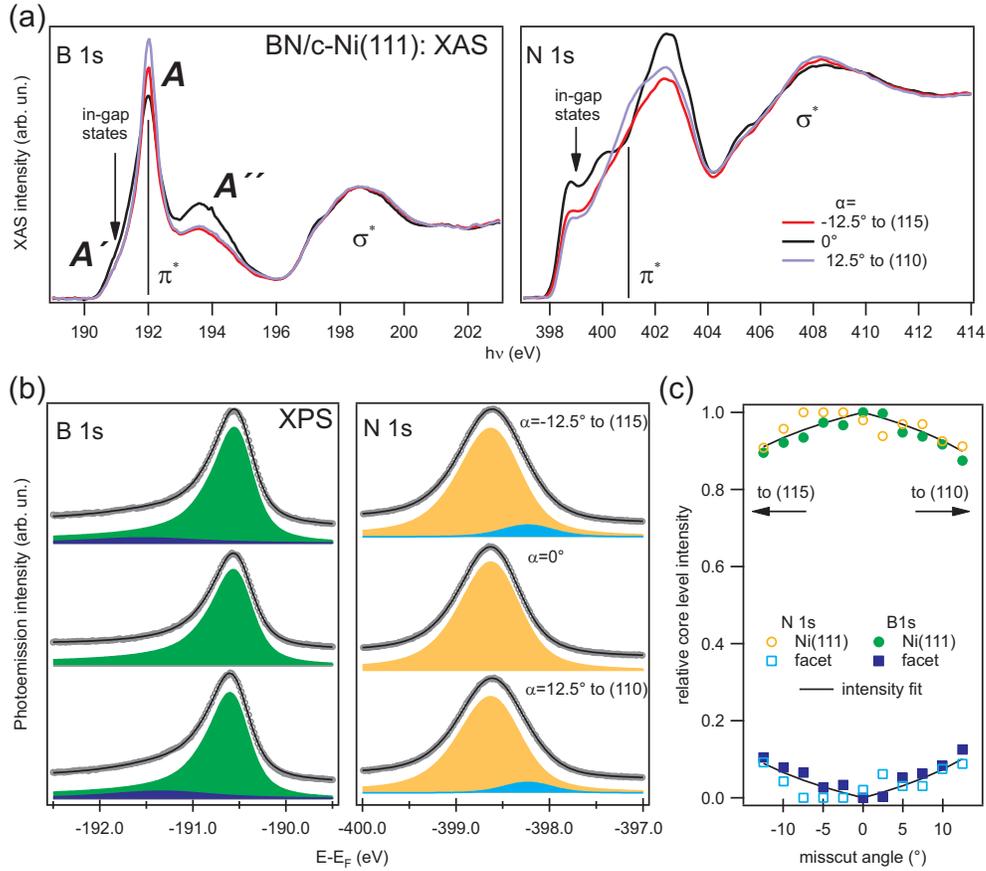}}
\caption{Near-edge X-ray absorption (NEXAFS) and X-ray Photoelectron Spectroscopy (XPS) of hBN/c-Ni(111): (a) XAS at the B and N $K$ adsorption edge and (b) XPS spectra for the B and N 1$s$ core levels, measured at $\alpha = \pm 12.5^{\circ}$ and 0$^{\circ}$ vicinal angles on the sample sketched in Fig.~\ref{STM_LEED}. NEXAFS spectra have been normalized using the $\sigma$ band, and $\pi^*$ features have been named following Ref. \cite{Preobrajenski2004_PRB}. In XPS spectra, each N and B 1$s$ emission has been fitted using the same pair of Doniach Sunjic lines at high and low binding energy. (c) Intensity of N and B 1$s$ XPS peaks (top, main lines, bottom, satellites) as a function of the vicinal angle. Experimental data are indicated by markers, whereas solid lines correspond to the expected variation assuming a satellite intensity that varies proportional to the area nominally covered by tilted facets.}
\label{XPS}
\end{figure}

Electronic states of the faceted hBN/c-Ni(111) system are investigated with Near Edge X-ray Absorption Spectroscopy (NEXAFS), X-ray Photoemission Spectroscopy (XPS) and Angle-Resolved Photoemission Spectroscopy (ARPES). In general, due to the relatively large width of the facets compared to atomic dimensions, superlattice or facet-boundary effects in the electronic structure are presumed to be very weak \cite{Corso09}. Therefore, one can reasonably expect a non-coherent superposition of two spectroscopic signals, arising from hBN-covered (111) terraces and (115)/(110) tilted facets, respectively. This is clearly observed in Fig.~\ref{XPS}, where we compare the boron and nitrogen NEXAFS $K$ adsorption edges and XPS core-level spectra taken at the (111) center and near the densely-stepped edges of the hBN/c-Ni(111) sample. The evolution of NEXAFS peaks from bulk to monolayer hBN on Ni(111) has already been characterized~\cite{Preobrajenski2004_PRB}. In the bulk hBN spectrum and at the B K-edge $\pi^{*}$ region, a single sharp \emph{A} transition at 192 eV is observed. At the hBN/Ni(111) interface, and due to chemical bonding with the Ni surface, the ``bulk'' \emph{A} peak broadens and decreases and two new peaks split off, reflecting the presence of in-gap metallic states (\emph{A'} edge) and the increasing weight of delocalized hBN conduction band states (\emph{A"} peak)~\cite{Preobrajenski2004_PRB,Laskowski2009_JPCM,SHIMOYAMA2009_JESR}. For hBN adsorption on the c-Ni(111) crystal in Fig.~\ref{XPS}(a) it is observed that \emph{A'} and \emph{A"} decrease, and the main \emph{A} peak increases and sharpens as we move from the center to the edges of the c-Ni(111) surface. This effect is observed at both, the B and N K-edge. We conclude that the overall interaction of hBN with the Ni surface reduces as we move to the faceted regions of the sample.
This could be contradictory with the fact that, compared to the close-packed (111) terraces, (115) or (110) facets exhibit a high number of Ni surface atoms with lower coordination (at step edges or in surface rows, respectively), which are expected to be highly reactive. However, the presence of an atomically rough interface may also result in a sizeable amount of N (B) atoms of the hBN layer that do not effectively contact the Ni substrate. In fact, assuming the planar hBN monolayer geometry suggested in the side view of Fig.~\ref{struc_model}, this is the case of the hBN/Ni(115) interface. Here, the strong corrugation of the stepped substrate leaves, at least, one out of three N (B) atoms far from the Ni(001) terrace. However, in the plane-parallel (5.15$\times$1) hBN/Ni(110) configuration assumed in Fig.~\ref{struc_model} the Ni(110) substrate corrugation is rather low. In this case, the presence of non-interacting N (B) atoms could only be explained by considering local buckling, that is, a non-rigid, atomically-rough hBN layer. This situation, though, has not been proved in the present STM images nor in LEED, and hence requires further validation.

The XPS experiments confirm the corrugated interface scenario in both Ni(115) and Ni(110) tilted facets, which leads to a significant decoupling of of the BN layer at the facets from the Ni substrate. As shown in Fig.~\ref{XPS}(b), N \emph{1s} and B 1\emph{s} peaks measured in XPS are broad due to the strong metallization of the hBN layer. At the center of the crystal, the N 1$s$ emission can be fitted by a single Doniach Sunjic (DS) line~\cite{Doniach1970} at $E$-$E_F$ = -398.63 eV. The B 1$s$ peak at $E$-$E_F$ = -190.56 eV is more asymmetric, due to the presence of high-binding energy shake-up excitations~\cite{PREOBRAJENSKI2005_SS}. At the faceted edges, both N 1$s$ and B 1$s$ exhibit a clear shift of the spectral weight to lower and higher binding energies, respectively. We have used a second DS fitting line to quantify this effect (blue shading) across the curved surface. The resulting satellite peaks are shifted with respect to the main line by $\sim$0.4 eV in N 1$s$ and by $\sim$-0.85 eV in B 1$s$, both, at A and B sides. The low/high binding energy shifts in N and B 1$s$ spectra are of the same order of those observed in hBN/Pt(111)~\cite{PREOBRAJENSKI2007_CPL}, reflecting the presence of less interacting, almost physisorbed N atoms, and slightly more interacting B atoms, adsorbed at positions other than hollow ones. The intensity of the satellite peak relative to the total N and B 1$s$ emission is plotted as a function of the vicinal angle in Fig.~\ref{XPS}(b). The relative intensity of the second peak increases proportional to the area of the facet. The latter is represented as solid lines that fit the data, with a proportionality constant of 1/5. Therefore, one out of five N atoms in both (115) and (110) facets are effectively decoupled from the Ni surface. This is in agreement with the rigid-hBN versus corrugated-substrate description of the hBN/Ni(115) facet of Fig.~\ref{struc_model}, while it suggests a non-planar conformation of the hBN layer covering the Ni(110) facet.

\begin{figure}[tb!]
\centerline{\includegraphics[width=140mm,angle=0,clip]{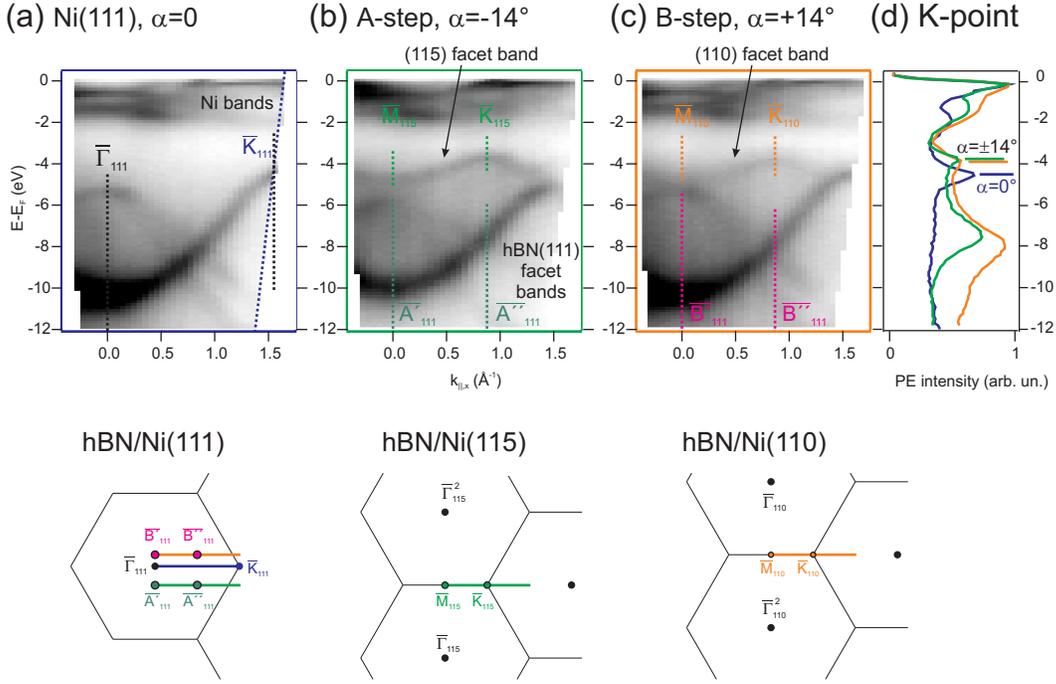}}
\caption{Angle resolved photoemission at the hBN/c-Ni(111) interface: Angular scans at sample (a) center, (b) A-side ($\alpha$ = 14$^{\circ}$), and (c) B-side ($\alpha$ = -14$^{\circ}$). The bottom panels indicate the hBN symmetry directions being probed for hBN/Ni(111) terraces [blue line in (a), green line in (b), and orange line in (c)], and hBN/Ni(115) [(b) scan] or hBN/Ni(110) [(c) scan] tilted facets. (d) Selected electron distribution curves from the angular scans in (a-c) (see vertical lines), at which the $\bar K$ point is probed in hBN/Ni(111) (blue), hBN/Ni(115) (green), and hBN/Ni(110) (orange). The photon energy is h$\nu$ = 40.8 eV (He II).}
\label{ARPES}
\end{figure}
The electronic band structure of the hBN/c-Ni(111) system is probed with ARPES using the He II excitation line at $h\nu$=40.8 eV. Characteristic photoemission intensity maps at the sample center and the densely stepped edges are shown in Fig.~\ref{ARPES}, together with drawings of the hBN reciprocal space that indicate the crystal directions being probed in each case. One observes that the photoemission signal at the double-phase, faceted areas of the surface is composed by the incoherent superposition of bands corresponding to hBN/Ni(111) terraces and hBN/Ni(115) (A side) or hBN/Ni(110) (B side) facets. A direct comparison between the image taken at the hBN/Ni(111) center in Fig.~\ref{ARPES} (a) with those corresponding to the faceted edges in Figs.~\ref{ARPES} (b) and (c) allows a rapid assignment of band features to each phase. In all cases, Ni substrate $d$ bands extend from the Fermi edge down to $E$-$E_F \sim$-2 eV, whereas bands below $\sim$-4 eV originate from the hBN monolayer. The measurement geometry was chosen to reach the $\bar K$-point, namely the valence band maximum (VBM) of hBN, through angular scans (azimuthal angle $\theta$, converted into the $k_{||}$ scale of Fig.~\ref{ARPES}), with the normal emission ($\theta$=0, tilt angle $\phi$=0) reference fixed along the [111] direction of bulk Ni. As shown in Fig.~\ref{ARPES} (a), in the hBN/Ni(111) center the $\phi$=0 scan directly delivers the $\bar \Gamma\bar K$ dispersion. At the sample edges, the emission from hBN covering Ni(115) and Ni(110) facets is significantly tilted from $\bar \Gamma$, although by selecting the correct $\phi$ angle in the two-dimensional detector we can track the dispersion along $\bar M\bar K$, and hence reach the $\bar K$ point of hBN on both Ni(115) [Fig.~\ref{ARPES} (b)] and Ni(110) [Fig.~\ref{ARPES} (c)] facets. For a better comparison of the VBM value in each case, we overlay and compare the electron distribution curves crossing the respective $\bar K$ point [marked as blue, green, and orange dotted lines on (a), (b) and (c) intensity plots] in Fig.~\ref{ARPES} (d). The $\bar K$ peak is found at -4.55 eV for hBN on Ni(111), in agreement with previous ARPES experiments \cite{Usachov2010_PRB}, and at -3.75 eV and -3.9 eV for the hBN/Ni(115) and hBN/Ni(110) interfaces. The latter agrees with the VBM value found in the hBN-covered Ni(110) crystal surface~\cite{Greber2006_EJSSN}.

The energy of the valence band maximum in the hBN monolayer has been shown to reflect the strength of its chemical interaction with the underlying substrate \cite{Nagashima1995_PRL,Grad2003_PRB,Roth2016_ACSNano}. Compared to the VBM of hBN/Ni(111), the higher value found in hBN/Ni(115) and hBN/Ni(110) facets indicates that, in average, the hBN-substrate interaction is lower at the two tilted facets. A similar effect has been observed already for graphene on faceted Ni~\cite{Grad2003_PRB,Usachov2007_SSP}. As discussed above, the presence of less-interacting hBN film to the underlying Ni may simply be explained by the existence of an inhomogeneous hBN/Ni contact at these two interfaces, with N and B atoms moving away from the on-top and hollow substrate positions, respectively, as detected in the XPS spectra for both hBN/Ni(115) and hBN/Ni(110).
Note, however, that strongly corrugated hBN layers grown in close-packed surfaces, such as the hBN nanomesh on Rh(111) \cite{Corso2004_Science,Roth2016_ACSNano}, posses N atoms with different substrate bonding configuration, and this leads to a $\sim 1.1$ eV $\pi$ band splitting. In the present case, a $\pi$ band splitting is not observed, although $\overline K$ point peaks in Fig.~\ref{ARPES} (d), particularly at the hBN/Ni(110) interface, appear broad, and hence a small splitting cannot be discarded.

\section{Conclusions}
hBN monolayer films can be successfully grown on a curved Ni(111) substrate, which exhibits vicinal surfaces with close-packed steps of variable density. Homogenous hBN coating of the entire curved surface is observed, although all its vicinal planes are transformed into faceted structures. These are made of hBN/Ni(111) terraces that alternate with tilted hBN/Ni(115) or hBN/Ni(110) nanostripes, respectively formed at A- and B-type vicinal surfaces, respectively.  X-ray absorption, X-ray and Angle-resolved photoemission spectroscopy agree to reveal that the hBN interaction with the substrate is stronger at Ni(111) planes, as compared to the (115) and (110) oriented facets. Our results suggest the use of Ni(115) and vicinal Ni(110) surface to achieve homogenous, monolayer thick hBN with weaker coupling with the Ni ferromagnetic substrate.
%The substrate that prior to the hBN adsorption presents well ordered single or double-layer high steps transforms then to a facet structure with two stable facets apart from the hBN/Ni(111) phase, namely hBN/Ni(115) on A-type and hBN/Ni(110) on B-type steps, respectively. The amount of the latter phases increases with increasing vicinal angle. The hBN films on the two side facets are less interacting with the substrate as observed by X-ray absorption and photoemission spectroscopy.
%On Ni(115) and Ni(110) the hBN film is not longer chemisorbed as for hBN/Ni(111).

%\section*{Acknowledgement}
\ack
We acknowledge financial support from the Deutsche Forschungsgemeinschaft DFG (Grants LA655-17/1 and LA655-19/1), the Spanish Ministry of Economy (Grants MAT-2016-78293-C6, MAT-2017-88374-P), the Basque Government (Grant IT-1255-19), the Helmholtz-Zentrum Berlin f\"ur Materialien und Energie for support within the bilateral Russian-German Laboratory program as well as from the joint project of the Russian Science Foundation (project 16-42-01093) and DFG (Grant No. LA655-17/1). D.Yu.U. acknowledges St. Petersburg State University for research Grant No. 11.65.42.2017 and Russian Foundation for Basic Research RFBR (Project No. 17-02-00427). Furthermore we thank Martina Corso for valuable scientific discussions, Dmitry Smirnov and Khadiza Ali for the support during the synchrotron and laboratory experiments, respectively.

\pagebreak

%\bibliographystyle{iop_articleTitles}
%\bibliography{dedicated}

\begin{thebibliography}{10}
\providecommand{\url}[1]{\texttt{#1}}
\providecommand{\urlprefix}{URL }
\providecommand{\eprint}[2][]{\url{#2}}

\bibitem{Pakdel2014_ChemSocRev}
Pakdel A, Bando Y and Golberg D 2014 Nano boron nitride flatland \emph{Chem.
  Soc. Rev.} \textbf{43} 934--959
%  \urlprefix\url{http://dx.doi.org/10.1039/C3CS60260E}

\bibitem{Rokuta1997_PRL}
Rokuta E, Hasegawa Y, Suzuki K, Gamou Y, Oshima C and Nagashima A 1997 Phonon
Dispersion of an Epitaxial Monolayer Film of Hexagonal Boron Nitride on
Ni(111) \emph{Phys. Rev. Lett.} \textbf{79} 4609--4612
%  \urlprefix\url{https://link.aps.org/doi/10.1103/PhysRevLett.79.4609}

\bibitem{Zunger1976_PRB}
Zunger A, Katzir A and Halperin A 1976 Optical properties of hexagonal boron
  nitride \emph{Phys. Rev. B} \textbf{13} 5560--5573
%  \urlprefix\url{https://link.aps.org/doi/10.1103/PhysRevB.13.5560}

\bibitem{SOLOZHENKO2001_JPhysChemSol}
Solozhenko V, Lazarenko A, Petitet J~P and Kanaev A 2001 Bandgap energy of
  graphite-like hexagonal boron nitride \emph{Journal of Physics and Chemistry
  of Solids} \textbf{62}(7) 1331 -- 1334
%  \urlprefix\url{http://www.sciencedirect.com/science/article/pii/S0022369701000300}

\bibitem{Watanabe2004_NatMat}
Watanabe K, Taniguchi T and Kanda H 2004 Direct-bandgap properties and evidence
  for ultraviolet lasing of hexagonal boron nitride single crystal \emph{Nature
  Materials} \textbf{3} 404
%\urlprefix\url{http://dx.doi.org/10.1038/nmat1134}

\bibitem{Watanabe2009_NatPhot}
Watanabe K, Taniguchi T, Niiyama T, Miya K and Taniguchi M 2009 Far-ultraviolet
  plane-emission handheld device based on hexagonal boron nitride \emph{Nature
  Photonics} \textbf{3} 591
%  \urlprefix\url{http://dx.doi.org/10.1038/nphoton.2009.167}

\bibitem{Rao2013_AngewChem}
Rao C~N~R, Ramakrishna~Matte H~S~S and Maitra U 2013 Graphene Analogues of
Inorganic Layered Materials \emph{Angewandte Chemie International Edition}
  \textbf{52}(50) 13162--13185
%  \eprint{https://onlinelibrary.wiley.com/doi/pdf/10.1002/anie.201301548}
%  \urlprefix\url{https://onlinelibrary.wiley.com/doi/abs/10.1002/anie.201301548}

\bibitem{Zhang2017_JMatChemC}
Zhang K, Feng Y, Wang F, Yang Z and Wang J 2017 Two dimensional hexagonal boron
  nitride (2D-hBN): synthesis{,} properties and applications \emph{J. Mater.
  Chem. C} \textbf{5} 11992--12022
%  \urlprefix\url{http://dx.doi.org/10.1039/C7TC04300G}

\bibitem{Usachov2010_PRB}
Usachov D, Adamchuk V~K, Haberer D, Gr\"uneis A, Sachdev H, Preobrajenski A~B,
  Laubschat C and Vyalikh D~V 2010 Quasifreestanding single-layer hexagonal
  boron nitride as a substrate for graphene synthesis \emph{Phys. Rev. B}
  \textbf{82} 075415
%  \urlprefix\url{https://link.aps.org/doi/10.1103/PhysRevB.82.075415}

\bibitem{Liu2011_NanoLett}
Liu Z, Song L, Zhao S, Huang J, Ma L, Zhang J, Lou J and Ajayan P~M 2011 Direct
Growth of Graphene/Hexagonal Boron Nitride Stacked Layers \emph{Nano Letters}
  \textbf{11}(5) 2032--2037 %pMID: 21488689
  %\urlprefix\url{https://doi.org/10.1021/nl200464j}

\bibitem{Britnell2012_Science}
Britnell L, Gorbachev R~V, Jalil R, Belle B~D, Schedin F, Mishchenko A,
  Georgiou T, Katsnelson M~I, Eaves L, Morozov S~V, Peres N~M~R, Leist J, Geim
  A~K, Novoselov K~S and Ponomarenko L~A 2012 Field-Effect Tunneling Transistor
  Based on Vertical Graphene Heterostructures \emph{Science} \textbf{335}(6071)
  947--950
  %\eprint{http://science.sciencemag.org/content/335/6071/947.full.pdf}
  %\urlprefix\url{http://science.sciencemag.org/content/335/6071/947}

\bibitem{Yang2013_NatMat}
Yang W, Chen G, Shi Z, Liu C~C, Zhang L, Xie G, Cheng M, Wang D, Yang R, Shi D,
  Watanabe K, Taniguchi T, Yao Y, Zhang Y and Zhang G 2013 Epitaxial growth of
  single-domain graphene on hexagonal boron nitride \emph{Nature Materials}
  \textbf{12} 792
  %\urlprefix\url{http://dx.doi.org/10.1038/nmat3695}

\bibitem{Kang2014_AdFuncMat}
Kang S~J, Lee G~H, Yu Y~J, Zhao Y, Kim B, Watanabe K, Taniguchi T, Hone J, Kim
  P and Nuckolls C 2014 Organic Field Effect Transistors Based on Graphene and
  Hexagonal Boron Nitride Heterostructures \emph{Advanced Functional Materials}
  \textbf{24}(32) 5157--5163
%  \eprint{https://onlinelibrary.wiley.com/doi/pdf/10.1002/adfm.201400348}
%  \urlprefix\url{https://onlinelibrary.wiley.com/doi/abs/10.1002/adfm.201400348}

\bibitem{Oh2016_NpgAsiaMat}
Oh H, Jo J, Tchoe Y, Yoon H, Hwi~Lee H, Kim S~S, Kim M, Sohn B~H and Yi G~C
  2016 Centimeter-sized epitaxial h-BN films \emph{Npg Asia Materials}
  \textbf{8} e330 %\urlprefix\url{https://doi.org/10.1038/am.2016.178}

\bibitem{Wang2015_AdvMat}
Wang H, Zhang X, Liu H, Yin Z, Meng J, Xia J, Meng X~M, Wu J and You J 2015
  Synthesis of large-sized single-crystal hexagonal boron nitride domains on
  nickel foils by ion beam sputtering deposition \emph{Advanced Materials}
  \textbf{27}(48) 8109--8115
%  \urlprefix\url{https://onlinelibrary.wiley.com/doi/abs/10.1002/adma.201504042}

\bibitem{Meng2017_Small}
Meng J, Zhang X, Wang Y, Yin Z, Liu H, Xia J, Wang H, You J, Jin P, Wang D and
  Meng X~M 2017 Aligned growth of millimeter-size hexagonal boron nitride
  single-crystal domains on epitaxial nickel thin film \emph{Small}
  \textbf{13}(18) 1604179
%  \urlprefix\url{https://onlinelibrary.wiley.com/doi/abs/10.1002/smll.201604179}

\bibitem{Li2014_Nanotechnology}
Li X, Yin J, Zhou J and Guo W 2014 Large area hexagonal boron nitride monolayer
  as efficient atomically thick insulating coating against friction and
  oxidation \emph{Nanotechnology} \textbf{25}(10) 105701
%  \urlprefix\url{http://stacks.iop.org/0957-4484/25/i=10/a=105701}

\bibitem{Caneva2017_ACS_ApplMat}
Caneva S, Martin M~B, D'Arsi{\'e} L, Aria A~I, Sezen H, Amati M, Gregoratti L,
  Sugime H, Esconjauregui S, Robertson J, Hofmann S and Weatherup R~S 2017 From
  growth surface to device interface: Preserving metallic Fe under monolayer
  hexagonal boron nitride \emph{ACS Applied Materials \& Interfaces}
  \textbf{9}(35) 29973--29981 %pMID: 28782356
%  \urlprefix\url{https://doi.org/10.1021/acsami.7b08717}

\bibitem{Spaeth2017_2DMat}
Sp{\"a}th F, Gebhardt J, D{\"u}ll F, Bauer U, Bachmann P, Gleichweit C,
  G{\"o}rling A, Steinr{\"u}ck H~P and Papp C 2017 Hydrogenation and hydrogen
  intercalation of hexagonal boron nitride on Ni(111): reactivity and
  electronic structure \emph{2D Materials} \textbf{4}(3) 035026
%  \urlprefix\url{http://stacks.iop.org/2053-1583/4/i=3/a=035026}

\bibitem{Schukin2013}
Shchukin V, Ledentsov N and Bimberg D 2013 \emph{Epitaxy of Nanostructures}
  NanoScience and Technology (Springer Berlin Heidelberg) ISBN 9783662070666
%  \urlprefix\url{https://books.google.es/books?id=zYbnCAAAQBAJ}

\bibitem{Markov2017}
Markov I~V 2017 \emph{Crystal Growth for Beginners} (WORLD SCIENTIFIC) 3rd edn.
%  \urlprefix\url{https://www.worldscientific.com/doi/abs/10.1142/10127}

\bibitem{Kuntze2002_APL}
Kuntze J, Mugarza A and Ortega J~E 2002 Ag-induced zero- and one-dimensional
  nanostructures on vicinal Si(111) \emph{Applied Physics Letters}
  \textbf{81}(13) 2463--2465
%\urlprefix\url{https://doi.org/10.1063/1.1509857}

\bibitem{Liu2003_Nanotech}
Liu B~Z and Nogami J 2003 Growth of parallel rare-earth silicide nanowire
  arrays on vicinal Si(001) \emph{Nanotechnology} \textbf{14}(8) 873
%  \urlprefix\url{http://stacks.iop.org/0957-4484/14/i=8/a=306}

\bibitem{Wang2011_ACSNano}
Wang J, Kaur I, Diaconescu B, Tang J~M, Miller G~P and Pohl K 2011 Highly
  ordered assembly of single-domain dichloropentacene over large areas on
  vicinal gold surfaces \emph{ACS Nano} \textbf{5}(3) 1792--1797

\bibitem{ROKUTA1999_SS}
Rokuta E, Hasegawa Y, Itoh A, Yamashita K, Tanaka T, Otani S and Oshima C 1999
  Vibrational spectra of the monolayer films of hexagonal boron nitride and
  graphite on faceted Ni(755) \emph{Surface Science} \textbf{427-428} 97 -- 101
%  \urlprefix\url{http://www.sciencedirect.com/science/article/pii/S0039602899002411}

\bibitem{Usachov2007_SSP}
Usachev D~Y, Shikin A~M, Varykhalov A~Y, Adamchuk V~K and Rader O 2007
  Angle-resolved photoelectron spectroscopy of geometrically nonuniform
  surfaces \emph{Phys. Solid State} \textbf{49} 949--957

\bibitem{Srut2013_Carbon}
{\^S}rut I, Trontl V~M, Pervan P and Kralj M 2013 Temperature dependence of
  graphene growth on a stepped Iridium surface \emph{Carbon} \textbf{56} 193 --
  200
%  \urlprefix\url{http://www.sciencedirect.com/science/article/pii/S000862231300047X}

\bibitem{Kajiwara2013_PRB}
Kajiwara T, Nakamori Y, Visikovskiy A, Iimori T, Komori F, Nakatsuji K, Mase K
  and Tanaka S 2013 Graphene nanoribbons on vicinal SiC surfaces by molecular
  beam epitaxy \emph{Phys. Rev. B} \textbf{87} 121407
%  \urlprefix\url{https://link.aps.org/doi/10.1103/PhysRevB.87.121407}

\bibitem{Celis2018_PRB}
Celis A, Nair M~N, Sicot M, Nicolas F, Kubsky S, Malterre D, Taleb-Ibrahimi A
  and Tejeda A 2018 Superlattice-induced minigaps in graphene band structure
  due to underlying one-dimensional nanostructuration \emph{Phys. Rev. B}
  \textbf{97} 195410
%  \urlprefix\url{https://link.aps.org/doi/10.1103/PhysRevB.97.195410}

\bibitem{Bachmann2003_SS}
Bachmann A, Speller S, Mugarza A and Ortega J 2003 Driving forces for
  Ag-induced periodic faceting of vicinal Cu(111) \emph{Surface Science}
  \textbf{526}(1) L143 -- L150
%  \urlprefix\url{http://www.sciencedirect.com/science/article/pii/S0039602802025748}

\bibitem{Riemann2005_PRB}
Riemann A, F\"olsch S and Rieder K~H 2005 Epitaxial growth of alkali halides on
  stepped metal surfaces \emph{Phys. Rev. B} \textbf{72} 125423

\bibitem{Johnson1994_PRL}
Johnson M~D, Orme C, Hunt A~W, Graff D, Sudijono J, Sander L~M and Orr B~G 1994
  Stable and unstable growth in molecular beam epitaxy \emph{Phys. Rev. Lett.}
  \textbf{72} 116--119
%  \urlprefix\url{https://link.aps.org/doi/10.1103/PhysRevLett.72.116}

\bibitem{Hornvonhoegen1999_SS}
von Hoegen M~H, zu~Heringdorf F~J~M, Hild R, Zahl P, Schmidt T and Bauer E 1999
  Au-induced giant faceting of vicinal Si(001) \emph{Surface Science}
  \textbf{433-435} 475 -- 480
%  \urlprefix\url{http://www.sciencedirect.com/science/article/pii/S0039602899004598}

\bibitem{Ortega2011_PRB}
Ortega J~E, Corso M, El-Fattah Z~M~A, Goiri E~A and Schiller F 2011 Interplay
  between structure and electronic states in step arrays explored with curved
  surfaces \emph{Phys. Rev. B} \textbf{83} 085411

\bibitem{Walter2015_NatComm}
Walter A~L, Schiller F, Corso M, Merte L~R, Bertram F, Lobo-Checa J, Shipilin
  M, Gustafson J, Lundgren E, Bri\'on-R\'{\i}os A~X, Cabrera-Sanfelix P,
  S\'anchez-Portal D and Ortega J~E 2015 X-ray photoemission analysis of clean
  and carbon monoxide-chemisorbed Platinum(111) stepped surfaces using a curved
  crystal \emph{Nature Communications} \textbf{6} 8903
%  \urlprefix\url{http://dx.doi.org/10.1038/ncomms9903}

\bibitem{Miccio2016_NL}
Miccio L~A, Setvin M, M{\"u}ller M, Abad{\'{\i}}a M, Piquero I, Lobo-Checa J,
  Schiller F, Rogero C, Schmid M, S{\'a}nchez-Portal D, Diebold U and Ortega
  J~E 2016 Interplay between steps and oxygen vacancies on curved TiO$_2$(110)
  \emph{Nano Letters} \textbf{16}(3) 2017--2022
%  \urlprefix\url{http://dx.doi.org/10.1021/acs.nanolett.5b05286}

\bibitem{Ortega2018_NJP}
Ortega J~E, Vasseur G, Piquero-Zulaica I, Matencio S, Valbuena M~A, Rault J~E,
  Schiller F, Corso M, Mugarza A and Lobo-Checa J 2018 Structure and electronic
  states of vicinal Ag(111) surfaces with densely kinked steps \emph{New
  Journal of Physics} \textbf{20}(7) 073010
%  \urlprefix\url{http://stacks.iop.org/1367-2630/20/i=7/a=073010}

\bibitem{Gamou1997_SR_RITU}
Gamou Y, Terai M, Nagashima A and Oshima C 1997 Atomic structural analysis of a
  monolayer epitaxial film of hexagonal boron nitride/Ni(111) studied by LEED
  intensity analysis \emph{Sci. Rep. RITU A} \textbf{44}(1) 211--214

\bibitem{Auwaerter1999_SS}
Auw{\"a}rter W, Kreutz T, Greber T and Osterwalder J 1999 XPD and STM
  investigation of hexagonal boron nitride on Ni(111) \emph{Surface Science}
  \textbf{429}(1) 229 -- 236
%  \urlprefix\url{http://www.sciencedirect.com/science/article/pii/S0039602899003817}

\bibitem{AUWARTER2004_ChemMat}
Auw{\"a}rter W, Suter H~U, Sachdev H and Greber T 2004 Synthesis of one
  monolayer of hexagonal boron nitride on Ni(111) from {\red{B}}-trichloroborazine
  (ClBNH)$_3$ \emph{Chemistry of Materials} \textbf{16}(2) 343--345
%  \urlprefix\url{https://doi.org/10.1021/cm034805s}

\bibitem{Lee2012_RCSAdv}
Lee Y~H, Liu K~K, Lu A~Y, Wu C~Y, Lin C~T, Zhang W, Su C~Y, Hsu C~L, Lin T~W,
  Wei K~H, Shi Y and Li L~J 2012 Growth selectivity of hexagonal-boron nitride
  layers on ni with various crystal orientations \emph{RSC Adv.} \textbf{2}
  111--115
%\urlprefix\url{http://dx.doi.org/10.1039/C1RA00703C}

\bibitem{Usachov2018_PRB}
Usachov D~Y, Tarasov A~V, Bokai K~A, Shevelev V~O, Vilkov O~Y, Petukhov A~E,
  Rybkin A~G, Ogorodnikov I~I, Kuznetsov M~V, Muntwiler M, Matsui F, Yashina
  L~V, Laubschat C and Vyalikh D~V 2018 Site- and spin-dependent coupling at
  the highly ordered $h$-BN/Co(0001) interface \emph{Phys. Rev. B} \textbf{98}
  195438
%  \urlprefix\url{https://link.aps.org/doi/10.1103/PhysRevB.98.195438}

\bibitem{Corso2004_Science}
Corso M, Auw{\"a}rter W, Muntwiler M, Tamai A, Greber T and Osterwalder J 2004
  Boron nitride nanomesh \emph{Science} \textbf{303}(5655) 217--220
%  \eprint{http://science.sciencemag.org/content/303/5655/217.full.pdf}
%  \urlprefix\url{http://science.sciencemag.org/content/303/5655/217}

\bibitem{PREOBRAJENSKI2007_CPL}
Preobrajenski A, Nesterov M, Ng M~L, Vinogradov A and M\aa{}rtensson N 2007
  Monolayer h-BN on lattice-mismatched metal surfaces: On the formation of the
  nanomesh \emph{Chemical Physics Letters} \textbf{446}(1) 119 -- 123
%  \urlprefix\url{http://www.sciencedirect.com/science/article/pii/S0009261407010998}

\bibitem{Goriachko2007_Langmuir}
Goriachko A, He, Knapp M, Over H, Corso M, Brugger T, Berner S, Osterwalder J
  and Greber T 2007 Self-assembly of a hexagonal boron nitride nanomesh on
  Ru(0001) \emph{Langmuir} \textbf{23}(6) 2928--2931 %pMID: 17286422
%  \urlprefix\url{https://doi.org/10.1021/la062990t}

\bibitem{Martoccia2008_PRL}
Martoccia D, Willmott P~R, Brugger T, Bj\"orck M, G\"unther S, Schlep\"utz C~M,
  Cervellino A, Pauli S~A, Patterson B~D, Marchini S, Wintterlin J, Moritz W
  and Greber T 2008 Graphene on Ru(0001): A
  $25\ifmmode\times\else\texttimes\fi{}25$ supercell \emph{Phys. Rev. Lett.}
  \textbf{101} 126102
%  \urlprefix\url{https://link.aps.org/doi/10.1103/PhysRevLett.101.126102}

\bibitem{Orlando2014_ACSNano}
Orlando F, Lacovig P, Omiciuolo L, Apostol N~G, Larciprete R, Baraldi A and
  Lizzit S 2014 Epitaxial growth of a single-domain hexagonal boron nitride
  monolayer \emph{ACS Nano} \textbf{8}(12) 12063--12070 %pMID: 25389799
%  \urlprefix\url{https://doi.org/10.1021/nn5058968}

\bibitem{Farwick2016_ACSNano}
Farwick~zum Hagen F~H, Zimmermann D~M, Silva C~C, Schlueter C, Atodiresei N,
  Jolie W, Mart\'{\i}nez-Galera A~J, Dombrowski D, Schr{\"o}der U~A, Will M,
  Lazi{\"c} P, Caciuc V, Bl{\"u}gel S, Lee T~L, Michely T and Busse C 2016
  Structure and growth of hexagonal boron nitride on Ir(111) \emph{ACS Nano}
  \textbf{10}(12) 11012--11026 %pMID: 28024332
%  \urlprefix\url{https://doi.org/10.1021/acsnano.6b05819}

\bibitem{PETROVIC2017_ASS}
Petrovi{\'c} M, Hagemann U, von Hoegen M~H and zu~Heringdorf F~J~M 2017
  Microanalysis of single-layer hexagonal boron nitride islands on Ir(111)
  \emph{Applied Surface Science} \textbf{420} 504 -- 510
%  \urlprefix\url{http://www.sciencedirect.com/science/article/pii/S0169433217314782}

\bibitem{Morscher2006_SS}
Morscher M, Corso M, Greber T and Osterwalder J 2006 Formation of single layer
  h-BN on Pd(111) \emph{Surface Science} \textbf{600}(16) 3280 -- 3284
%  \urlprefix\url{http://www.sciencedirect.com/science/article/pii/S0039602806007266}

\bibitem{PAFFETT1990_SS}
Paffett M, Simonson R, Papin P and Paine R 1990 Borazine adsorption and
  decomposition at Pt(111) and Ru(001) surfaces \emph{Surface Science}
  \textbf{232}(3) 286 -- 296
%  \urlprefix\url{http://www.sciencedirect.com/science/article/pii/003960289090121N}

\bibitem{PREOBRAJENSKI2007_PRB}
Preobrajenski A~B, Vinogradov A~S, Ng M~L, \ifmmode~\acute{C}\else
  \'{C}\fi{}avar E, Westerstr\"om R, Mikkelsen A, Lundgren E and M\aa{}rtensson
  N 2007 Influence of chemical interaction at the lattice-mismatched
  $h$-BN/Rh(111) and $h$-BN/Pt(111) interfaces on the overlayer morphology
  \emph{Phys. Rev. B} \textbf{75} 245412
%  \urlprefix\url{https://link.aps.org/doi/10.1103/PhysRevB.75.245412}

\bibitem{CAVAR2008_SS}
{\'C}avar E, Westerstr{\"o}m R, Mikkelsen A, Lundgren E, Vinogradov A, Ng M~L,
  Preobrajenski A, Zakharov A and M\aa{}rtensson N 2008 A single h-BN layer on
  Pt(111) \emph{Surface Science} \textbf{602}(9) 1722 -- 1726
%  \urlprefix\url{http://www.sciencedirect.com/science/article/pii/S0039602808001672}

\bibitem{PREOBRAJENSKI2005_SS}
Preobrajenski A, Vinogradov A and M\aa{}rtensson N 2005 Monolayer of h-BN
  chemisorbed on Cu(111) and Ni(111): The role of the transition metal 3d
  states \emph{Surface Science} \textbf{582}(1) 21 -- 30
%  \urlprefix\url{http://www.sciencedirect.com/science/article/pii/S0039602805002426}

\bibitem{Joshi2012_NanoLett}
Joshi S, Ecija D, Koitz R, Iannuzzi M, Seitsonen A~P, Hutter J, Sachdev H,
  Vijayaraghavan S, Bischoff F, Seufert K, Barth J~V and Auw{\"a}rter W 2012
  Boron nitride on Cu(111): An electronically corrugated monolayer \emph{Nano
  Letters} \textbf{12}(11) 5821--5828 %pMID: 23083003
%  \urlprefix\url{https://doi.org/10.1021/nl303170m}

\bibitem{Mueller2010_PRB}
M\"uller F, H\"ufner S, Sachdev H, Laskowski R, Blaha P and Schwarz K 2010
  Epitaxial growth of hexagonal boron nitride on Ag(111) \emph{Phys. Rev. B}
  \textbf{82} 113406
%  \urlprefix\url{https://link.aps.org/doi/10.1103/PhysRevB.82.113406}

\bibitem{Greber2006_EJSSN}
Greber T, Brandenberger L, Corso M, Tamai A and Osterwalder J 2006 Single layer
  hexagonal boron nitride films on Ni(110) \emph{e-Journal of Surface Science
  and Nanotechnology} \textbf{4} 410--413

\bibitem{Vinogradov2012_Langmuir}
Vinogradov N~A, Zakharov A~A, Ng M~L, Mikkelsen A, Lundgren E, M\aa{}rtensson N
  and Preobrajenski A~B 2012 One-dimensional corrugation of the h-BN monolayer
  on Fe(110) \emph{Langmuir} \textbf{28}(3) 1775--1781 pMID: 22185488
%  \urlprefix\url{https://doi.org/10.1021/la2035642}

\bibitem{Mueller2008_SS}
M{\"u}ller F, H{\"u}fner S and Sachdev H 2008 One-dimensional structure of
  boron nitride on Chromium (110) - a study of the growth of boron nitride by
  chemical vapour deposition of borazine \emph{Surface Science}
  \textbf{602}(22) 3467 -- 3476
%  \urlprefix\url{http://www.sciencedirect.com/science/article/pii/S0039602808004287}

\bibitem{HERRMANN2018_SS}
Herrmann C, Omelchenko P and Kavanagh K~L 2018 Growth of h-BN on Copper (110)
  in a LEEM \emph{Surface Science} \textbf{669} 133 -- 139
%  \urlprefix\url{http://www.sciencedirect.com/science/article/pii/S0039602817306957}

\bibitem{CORSO2005_SS}
Corso M, Greber T and Osterwalder J 2005 h-BN on Pd(110): a tunable system for
  self-assembled nanostructures? \emph{Surface Science} \textbf{577}(2) L78 --
  L84
%  \urlprefix\url{http://www.sciencedirect.com/science/article/pii/S0039602805000531}

\bibitem{Allan2007_NanoScReLe}
Allan M~P, Berner S, Corso M, Greber T and Osterwalder J 2007 Tunable
  self-assembly of one-dimensional nanostructures with orthogonal directions
  \emph{Nanoscale Research Letters} \textbf{2}(2) 94
%  \urlprefix\url{https://doi.org/10.1007/s11671-006-9036-2}

\bibitem{Ilyn2016_JPCC}
Ilyn M, Maga{\~n}a A, Walter A~L, Lobo-Checa J, de~Oteyza D~G, Schiller F and
  Ortega J~E 2017 Step-doubling at vicinal Ni(111) surfaces investigated with a
  curved crystal \emph{The Journal of Physical Chemistry C} \textbf{121}(7)
  3880--3886

\bibitem{Lynch1966_JCP}
Lynch R~W and Drickamer H~G 1966 Effect of high pressure on the lattice
  parameters of diamond, graphite, and hexagonal boron nitride \emph{The
  Journal of Chemical Physics} \textbf{44}(1) 181--184
 % \eprint{https://doi.org/10.1063/1.1726442}
%  \urlprefix\url{https://doi.org/10.1063/1.1726442}

\bibitem{BANDYOPADHYAY1977_Cryo}
Bandyopadhyay J and Gupta K 1977 Low temperature lattice parameter of Nickel
  and some Nickel-Cobalt alloys and Gr\"uneisen parameter of Nickel
  \emph{Cryogenics} \textbf{17}(6) 345 -- 347
%  \urlprefix\url{http://www.sciencedirect.com/science/article/pii/0011227577901308}

\bibitem{Grad2003_PRB}
Grad G~B, Blaha P, Schwarz K, Auw{\"a}rter W and Greber T 2003 Density
  functional theory investigation of the geometric and spintronic structure of
  h-BN/Ni(111) in view of Photoemission and STM experiments \emph{Phys. Rev. B}
  \textbf{68} 085404
%  \urlprefix\url{https://link.aps.org/doi/10.1103/PhysRevB.68.085404}

\bibitem{Sun2014_JAP}
Sun X, Pratt A, Li Z~Y, Ohtomo M, Sakai S and Yamauchi Y 2014 The adsorption of
  h-BN monolayer on the Ni(111) surface studied by density functional theory
  calculations with a semiempirical long-range dispersion correction
  \emph{Journal of Applied Physics} \textbf{115}(17) 17C117
%  \urlprefix\url{https://doi.org/10.1063/1.4866237}

\bibitem{EBNONNASIR2015_SRL}
Ebnonnasir A, Kodambaka S and Ciobanu C~V 2015 Strongly and weakly interacting
  configurations of hexagonal boron nitride on nickel \emph{Surface Review and
  Letters} \textbf{22}(06) 1550078
%  \urlprefix\url{https://doi.org/10.1142/S0218625X1550078X}

\bibitem{note1}
This finding confirms an earlier analysis of the hBN/Ni(755) interface, where
  (111) terraces and 45\° inclined facets were observed~\cite{ROKUTA1999_SS}.

\bibitem{note2}
Note, however, that in the kinematic approach double atomic steps leads to a
  (2$\times$1) superstructure with equally intense spots. In
  Fig.~\ref{STM_LEED} (b) we observe an alternating intensity of the
  (2$\times$1) spots, which could be due to multiple scattering in the
  step-doubled lattice. Also the presence of some minor $1 \times 1$
  (single-atom high steps) domains could explain the observed (2$\times$1)
  intensity variation.

\bibitem{Che2005_PRB}
Che J~G and Cheng H~P 2005 First-principles investigation of a monolayer of
  ${\mathrm{C}}_{60}$ on $h$-BN/Ni(111) \emph{Phys. Rev. B} \textbf{72} 115436
%  \urlprefix\url{https://link.aps.org/doi/10.1103/PhysRevB.72.115436}

\bibitem{Huda2006_PRB}
Huda M~N and Kleinman L 2006 $h$-$\mathrm{BN}$ monolayer adsorption on the
  $\mathrm{Ni}\phantom{\rule{0.3em}{0ex}}(111)$ surface: A density functional
  study \emph{Phys. Rev. B} \textbf{74} 075418

\bibitem{Joshi2013_PRB}
Joshi N and Ghosh P 2013 Substrate-induced changes in the magnetic and
  electronic properties of hexagonal boron nitride \emph{Phys. Rev. B}
  \textbf{87} 235440
%  \urlprefix\url{https://link.aps.org/doi/10.1103/PhysRevB.87.235440}

\bibitem{AUWARTER2003_SS}
Auw{\"a}rter W, Muntwiler M, Osterwalder J and Greber T 2003 Defect lines and
  two-domain structure of hexagonal boron nitride films on Ni(111)
  \emph{Surface Science} \textbf{545}(1) L735 -- L740
%  \urlprefix\url{http://www.sciencedirect.com/science/article/pii/S0039602803010902}

\bibitem{Osterwalder2003_EJSSN}
Osterwalder J, Auw{\"a}rter W, Muntwiler M and Greber T 2003 {Growth
  morphologies and defect structure in hexagonal boron nitride films on
  Ni(111): A Combined STM and XPD Study} \emph{e-Journal of Surface Science and
  Nanotechnology} \textbf{1} 124--129

\bibitem{Preobrajenski2004_PRB}
Preobrajenski A~B, Vinogradov A~S and M\aa{}rtensson N 2004 Ni $3d$-BN $\pi$
  hybridization at the $h$-BN/Ni(111) interface observed with core-level
  spectroscopies \emph{Phys. Rev. B} \textbf{70} 165404
%  \urlprefix\url{https://link.aps.org/doi/10.1103/PhysRevB.70.165404}

\bibitem{Corso09}
Corso M, Schiller F, Fern{\'a}ndez L, Cord{\'o}n J and Ortega J~E 2009
  Electronic states in faceted Au(111) studied with curved crystal surfaces
  \emph{Journal of Physics: Condensed Matter} \textbf{21}(35) 353001
%  \urlprefix\url{http://stacks.iop.org/0953-8984/21/i=35/a=353001}

\bibitem{Laskowski2009_JPCM}
Laskowski R, Gallauner T, Blaha P and Schwarz K 2009 Density functional theory
  simulations of B K and N K Nexafs spectra of h-BN/transition metal(111)
  interfaces \emph{Journal of Physics: Condensed Matter} \textbf{21}(10) 104210
%  \urlprefix\url{http://stacks.iop.org/0953-8984/21/i=10/a=104210}

\bibitem{SHIMOYAMA2009_JESR}
Shimoyama I, Baba Y, Sekiguchi T and Nath K 2009 A theoretical interpretation
  of near edge x-ray absorption fine structure of hexagonal boron nitride
  monolayer on Ni(111) \emph{Journal of Electron Spectroscopy and Related
  Phenomena} \textbf{175}(1) 6 -- 13
%  \urlprefix\url{http://www.sciencedirect.com/science/article/pii/S0368204809001649}

\bibitem{Doniach1970}
Doniach S and \v{S}unjic M 1970 Many-electron singularity in x-ray
  photoemission and x-ray line spectra from metals \emph{Journal of Physics C:
  Solid State Physics} \textbf{3}(2) 285
%  \urlprefix\url{http://stacks.iop.org/0022-3719/3/i=2/a=010}

\bibitem{Nagashima1995_PRL}
Nagashima A, Tejima N, Gamou Y, Kawai T and Oshima C 1995 Electronic structure
  of monolayer hexagonal boron nitride physisorbed on metal surfaces
  \emph{Phys. Rev. Lett.} \textbf{75} 3918--3921
%  \urlprefix\url{https://link.aps.org/doi/10.1103/PhysRevLett.75.3918}

\bibitem{Roth2016_ACSNano}
Roth S, Greber T and Osterwalder J 2016 Some like it flat: Decoupled h-BN
  monolayer substrates for aligned graphene growth \emph{ACS Nano}
  \textbf{10}(12) 11187--11195

\end{thebibliography}

\end{document}